\documentclass[sigconf]{acmart}
\copyrightyear{2026}
\acmYear{2026}
\setcopyright{cc}
\setcctype{by}
\acmConference[CHI EA '26]{Extended Abstracts of the 2026 CHI Conference on Human Factors in Computing Systems}{April 13--17, 2026}{Barcelona, Spain}
\acmBooktitle{Extended Abstracts of the 2026 CHI Conference on Human Factors in Computing Systems (CHI EA '26), April 13--17, 2026, Barcelona, Spain}
\acmDOI{10.1145/3772363.3799031}
\acmISBN{979-8-4007-2281-3/2026/04}

\usepackage{multirow}
\usepackage{makecell}
\usepackage{microtype}
\usepackage{enumitem}
\usepackage{etoolbox}
\usepackage{xcolor}

\newcolumntype{L}[1]{>{\raggedright\arraybackslash}p{#1}}
\makeatletter
\newcommand{\squote}{\@ifnextchar[{\squote@opt}{\squote@noopt}}
\newcommand{\squote@opt}[2][]{%
  \textit{``#2''}~(\textit{#1})%
}
\newcommand{\squote@noopt}[1]{%
  \textit{``#1''}%
}
\makeatother

\makeatletter
\newcommand{\lquote}{\@ifnextchar[{\lquote@opt}{\lquote@noopt}}
\newcommand{\lquote@opt}[2][]{%
  \begin{quote}
  ``#2''---\textit{#1}%
  \end{quote}
}
\newcommand{\lquote@noopt}[1]{%
  \begin{quote}
  ``#1''%
  \end{quote}
}
\makeatother

\makeatletter
\newcommand{\optional}[2][0]{%
  \def\optarg{#1}%
  \ifnum\optarg=1\relax
    #2%
  \fi
}
\makeatother
\begin{document}

% \title{Exploring the Role of Comments Before, During, and After Task-Learning with Videos}
% \title{From Search to Reflect: Exploring the Role of Comments for Task-Learning with Videos}
\title{Exploring the Role of User Comments Throughout the Stages of Video-Based Task-Learning}

% Yotam*,
% Nayoung*,
% Zhongyi,
% Igarashi

\settopmatter{authorsperrow=4}

\author{Nayoung Kim}
\authornote{Both authors contributed equally to this work.}
\affiliation{
  \institution{KAIST}
  \city{Daejeon}
  \country{South Korea}
}
\email{skdud727@kaist.ac.kr}

\author{Yotam Sechayk}
\authornotemark[1]
%\authornote{Both authors contributed equally to this work.}
\affiliation{
  \institution{The University of Tokyo}
  \city{Tokyo}
  \country{Japan}
}
\email{sechayk-yotam@g.ecc.u-tokyo.ac.jp}

\author{Zhongyi Zhou}
\affiliation{
  \institution{Google}
  \city{Tokyo}
  \country{Japan}
}
\email{zhongyizhou@google.com}

\author{Takeo Igarashi}
\affiliation{
  \institution{The University of Tokyo}
  \city{Tokyo}
  \country{Japan}
}
\email{takeo@acm.org}

%\renewcommand{\shortauthors}{Author, et al.}

\begin{abstract}
    Learning tasks through videos is a dynamic way to acquire skills by witnessing entire processes. However, compared to in-person demonstrations, videos may omit tacit knowledge, including subtle details and contextual nuances. Users' unique circumstances, like missing ingredients in a recipe, may also require adaptation beyond the video content. To fill these gaps, many users turn to the comment section, seeking additional guidance and interactions with creators or peers to personalize their experience. Despite their importance, there is limited understanding of how users engage with and apply comments in task-learning scenarios. In our study, we explore the role of comments in video-based task-learning through interviews with 14 users, and co-watching sessions with eight. Our findings show that while comments are critical for learning, they are poorly integrated into all stages of the learning process. Based on our findings, we outline design opportunities to better utilize comments in video-based task-learning.
\end{abstract}




\begin{teaserfigure}
    \centering
    \includegraphics[width=0.8\textwidth]{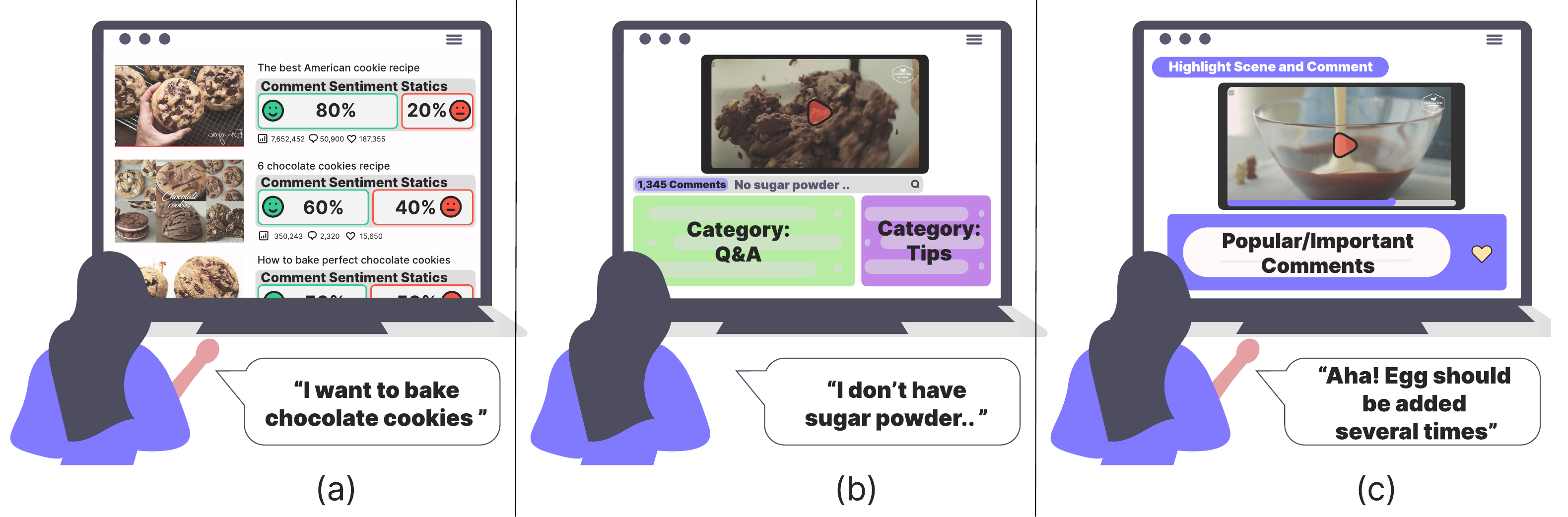}
    \caption{Design opportunities for utilizing comments in video-based task-learning, based on an exploration of current usage of comments by viewers. Taking advantage of (a) sentiment analysis from comments, (b) categorization of comments, and (c) reflecting on task-learning videos through observations from comments.}
    \Description{The figure consists of three panels, illustrating a user interface for video-based learning with a focus on comment interaction and sentiment analysis. In panel (a), a computer screen displays a list of cookie-baking videos with thumbnails, titles, and metadata such as view counts, likes, and comment numbers. Each video features "Comment Sentiment Statistics," shown as percentages with happy and frowning face icons (e.g., 80\% positive and 20\% negative). A user, represented as a purple silhouette with long hair, points at the screen with a speech bubble saying, "I want to bake chocolate cookies." In panel (b), the interface shifts to a single video paused on a close-up of cookie dough, with comments organized into categories like “Q\&A” (green) and “Tips” (purple). Below the screen, the user expresses a concern with a speech bubble saying, "I don’t have sugar powder..." In panel (c), the video player highlights a specific scene about adding eggs, labeled “Highlight Scene and Comment.” Below, a section titled “Popular/Important Comments” appears, while the user reacts with a speech bubble saying, "Aha! Egg should be added several times." Across all panels, the user actively engages with the content, reflecting emotions and learning progress through their comments.}
    \label{fig:teaser}
\end{teaserfigure}


\maketitle

\section{Introduction}
\label{sec:introduction}

Learning tasks through videos has become an increasingly common practice, offering a dynamic and visual medium for acquiring new skills and knowledge~\cite{shen2023LearnerEngagement,manca2018SocialNetwork,zhao2020VideoQuestion,dubovi2020EmpiricalAnalysis}.
Through task-learning videos, viewers can learn a wide range of skills, from music~\cite{waldron2013UsergeneratedContent} or cooking~\cite{oguz2022ChopChange}, to programming~\cite{khandwala2018CodemotionExpanding} or math~\cite{yang2020LearningAlone}.
Unlike other resources such as written media, videos function as \textit{demonstrations}, which can provide additional tacit knowledge---unspoken, intuitive, or experiential insights~\cite{oomori2024SkillsInterpreterCase}.
In-person demonstrations provide viewers autonomy to freely determine points of interest; however, videos have a predefined point-of-view, which limits opportunities for tacit knowledge acquisition~\cite{felice2021SocialInteraction,john2016CollationTwo,ramlogan2014ComparisonTwo,defelice2022LearningOthers,solomon2004RandomizedTrial}.
Moreover, videos may lack instructions that directly align with the needs of each viewer.  For example, in cooking videos, viewers might not have access to all ingredients or tools the recipe calls for.

To bridge these gaps, viewers frequently turn to the comment section as a supplementary resource for task-learning~\cite{thelwall2012CommentingYouTube,lee2015UsingTimeanchored,lee2017MakingSense}.
Viewers contribute to this space by asking clarifying questions, offering explanations, and sharing their goals or struggles, collectively shaping the comment section into a collaborative sensemaking site~\cite{writingcomments}. These interactions facilitate a shared learning experience, enabling viewers to personalize their learning journey while benefiting from the collective wisdom of the community~\cite{lim2021SharingLearning,wang2023LetsPlay,bulbul2021SocialLearning,dubovi2020EmpiricalAnalysis}.
Prior work has primarily examined these visible interactions through comment-based analyses~\cite{dubovi2020EmpiricalAnalysis,yang2020LearningAlone,lim2021SharingLearning,nguyen2023HowCivil,zeng2024DanmakubasedAutomatic}. However, despite recognizing comments as a collaborative resource, we still have limited understanding of how viewers read, interpret, and strategically integrate comments into their learning process. Without this understanding, it remains unclear how comments shape learning outcomes or how they might be better designed to support task-learning. This gap is particularly evident in task-learning videos, where user interaction through comments has received far less attention than in educational videos~\cite{lee2017MakingSense,sung2016ToPINVisual,dubovi2020EmpiricalAnalysis,thelwall2012CommentingYouTube}.

%To bridge these gaps, viewers frequently turn to the comment section as a supplementary resource for task-learning~\cite{thelwall2012CommentingYouTube,lee2015UsingTimeanchored,lee2017MakingSense}.
%Comments provide a collaborative space where viewers can seek guidance, clarify uncertainties, and interact with both video creators and other learners~\cite{tanprasert2023ScriptedVicarious,wu2019DanmakuNew,zeng2024DanmakubasedAutomatic}.
%These interactions facilitate a shared learning experience, enabling viewers to personalize their learning journey while benefiting from the collective wisdom of the community~\cite{lim2021SharingLearning,wang2023LetsPlay,bulbul2021SocialLearning,dubovi2020EmpiricalAnalysis}.
%Several works explore viewer interaction through comment-based analysis~\cite{dubovi2020EmpiricalAnalysis,yang2020LearningAlone,lim2021SharingLearning,nguyen2023HowCivil,zeng2024DanmakubasedAutomatic}, however, we have limited understanding on the experience of users when interacting with video comments for task-learning.
% Moreover, this gap is particularly evident in task-learning videos, where user interaction through comments has received far less attention than in educational videos~\cite{lee2017MakingSense,sung2016ToPINVisual,dubovi2020EmpiricalAnalysis,thelwall2012CommentingYouTube}.

In this work, we investigate the usage of comments in video-based task-learning. 
Specifically, we conducted exploratory interviews (N=14) and co-watching sessions (N=8) to identify patterns in how participants use comments for learning.
Our findings reveal that users refer to comments throughout their learning process---when searching for videos, actively learning, and when reflecting on the content---however, the integration of comments as a learning tool remains limited. 
This highlights an opportunity to better incorporate comments into the task-learning process, transforming them from a purely social interaction tool into a valuable extension of the learning environment.
Based on our findings, we discuss potential opportunities to further integrate comments into the task-learning process.

% Our contributions are as follows:
% \begin{enumerate}
%     \item Exploratory interviews and co-watching sessions on the role of comments in task-learning.
%     \item Suggestion of design opportunities for future task-learning systems integrating video comments.
% \end{enumerate}

\section{Background}
\label{sec:background}

In-person demonstrations allow for student-to-student or student-to-instructor interactions; however, videos are pre-recorded streams of information, which leads users to turn to the comment section~\cite{gao2024EvaluatingLearners}.
Using comments can cultivate a sense of community~\cite{gao2024EvaluatingLearners,yang2024CombiningDanmaku,bulbul2021SocialLearning}, which both encourages and supports learners~\cite{huang2024ExaminingRole,du2009FeltMore}.
In task-learning videos, comments can lead to a shared learning experience with collaboratively constructed knowledge~\cite{dubovi2020EmpiricalAnalysis,lim2021SharingLearning,yang2020LearningAlone,wang2023LetsPlay,thompsonHowKhan,yang2024CombiningDanmaku}.
Both positive and negative comments can influence viewers~\cite{moller2023OnlineSocial,kramer2021FeelWhat,towne2017ConflictComments}.
Positive comments can improve motivation and engagement~\cite{huang2024ExaminingRole}, reduce anxiety~\cite{deloatch2017NeedYour}, and enhance the learning~\cite{benson2015CommentingLearn,leng2016IdentifyingPotential}.
Negative comments, like critical feedback, can lower perceived quality but still foster learning~\cite{towne2017ConflictComments}.
While most works explore the nature of viewer interactions through analysis of existing comments~\cite{dubovi2020EmpiricalAnalysis,yang2020LearningAlone,lim2021SharingLearning,nguyen2023HowCivil,zeng2024DanmakubasedAutomatic}, little is known on how viewers use comments in relation to the learning process (i.e., before, during, and after video watching).
This work explores how comments are being used in practice by viewers through interviews and co-watching sessions. 

Several studies have explored leveraging comments to enhance the learning.
\citet{bunt2014TaggedCommentsPromoting} enabled users to tag comments on written guides to support filtering and navigation.
In video-based learning, systems such as TrACE~\cite{dorn2015PilotingTrACE} and ToPIN~\cite{sung2016ToPINVisual} introduced time-anchored comments and visualizations to support comment exploration and navigation.
Other works examined comments as a feedback mechanism for instructors~\cite{astudillo2024HowCould}, or introduced fictional comments to promote social and vicarious learning~\cite{tanprasert2023ScriptedVicarious}.
Together, these efforts demonstrate the potential of comments to support sensemaking, reflection, and interaction in video-based learning.
In this work, we discuss design opportunities to further utilize user comments throughout the learning process.

\section{Exploratory Study}
\label{sec:exploratory}
Drawing on insights from interviews and co-watching sessions, this section examines the role of comments across different phases of the learning process, exploring when and how they become an essential tool for task-learning.

\subsection{Methodology}
\subsubsection{Participants}
We recruited 14 participants (7 male, 7 female), aged 20 to 26 years (M = 23.5, SD = 1.6), all affiliated with our university.
Participants were recruited through peer networks and had varied interests and prior experiences with video-based task-learning.
All participants completed interviews, and 8 additionally participated in co-watching sessions based on availability.
Our goal was not statistical generalization but an in-depth understanding of users’ experiences with comments.
Participant demographics are summarized in Table~\ref{tab:participants}, and ethical considerations are described in Appendix~\ref{apx:ethical}.

\subsubsection{Procedure}
Interviews lasted approximately 30 minutes, with co-watching sessions taking an additional 30 minutes to complete. We explored when, how, and why participants use comments while learning tasks through videos.
The interview protocol was developed by two of the authors following \citet{lazar2017ResearchMethods}.
During the co-watching sessions, participants selected a task and video of personal interest, and interacted with comments on their preferred platform to reflect realistic learning situations. All participants chose YouTube~\cite{youtube} as their platform of choice during this study.
The study was conducted remotely via Zoom~\footnote{https://www.zoom.com/}, with sessions recorded, transcribed locally using OpenAI Whisper~\cite{radford2022whisper}, and analyzed using reflexive thematic analysis~\cite{braun2019ReflectingReflexive}, with two researchers iteratively coding and discussing themes related to comment use across different stages of the task-learning process. 
% Throughout this section, participants’ backgrounds refer to task-related interests and learning contexts, not demographic characteristics.
The interview protocol is included in Appendix~\ref{apx:exploratory-questions}.

\begin{table*}[h]
    \caption{Participant demographics. † indicates participation in co-watching sessions.}
    \label{tab:participants}
    \centering
    \small

\begin{tabular}{llllll}
\toprule
PID  & Age & Gender & Video Learning Frequency         & Preferred Device & \begin{tabular}[c]{@{}c@{}}Co-watching Subject\end{tabular} \\ \midrule
P1   & 25  & Male   & Sometimes (once or twice a week) & Smartphone       & ---                                                              \\
P2†  & 24  & Female & Often (a few times a week)       & Smartphone       & Sewing                                                           \\
P3   & 26  & Male   & Sometimes (once or twice a week) & Smartphone       & ---                                                              \\
P4   & 26  & Male   & Often (a few times a week)       & Smartphone       & ---                                                              \\
P5   & 23  & Female & Sometimes (once or twice a week) & Smartphone       & ---                                                              \\
P6†  & 23  & Female & Sometimes (once or twice a week) & Smartphone       & Baking                                                           \\
P7   & 23  & Male   & Sometimes (once or twice a week) & Smartphone       & ---                                                              \\
P8   & 23  & Female & Often (a few times a week)       & Smartphone       & ---                                                              \\
P9†  & 20  & Female & Sometimes (once or twice a week) & Desktop          & Hair Styling                                                     \\
P10† & 23  & Female & Frequently (almost daily)        & Smartphone       & Cooking                                                          \\
P11† & 23  & Female & Often (a few times a week)       & Smartphone       & Hair Styling                                                     \\
P12† & 26  & Male   & Often (a few times a week)       & Smartphone       & Sewing                                                           \\
P13† & 23  & Male   & Frequently (almost daily)        & Smartphone       & Hair Styling                                                     \\
P14† & 22  & Male   & Sometimes (once or twice a week) & Smartphone       & Card Shuffling                                                            \\ \bottomrule
\end{tabular}

\end{table*}

\subsection{Findings}
\label{sec:findings}
Participants highlighted the benefits of videos for task-learning across various topics, such cooking/baking (P1-P11) or exercise (P1, P2, P5, P6, P8, P9, P11-P14) among others.
% Cooking (P1-P11), exercise (P1, P2, P5, P6, P8, P9, P11-P14), hair styling (P2, P7-P13), fashion (P2, P5, P6, P9, P10, P12), programming (P2, P5-P8, P12), and makeup application (P5, P6, P8, P10).
As P12 noted, compared to written media \squote{videos allow me to observe the in-between process as well.}
Participants found videos more engaging and easier to follow and navigate, particularly for physical tasks, emphasizing the benefits of visual navigation (P2, P5, P7, P9, P10, P11, P13). 
We detail how participants interact with comments, and how they incorporate them into their learning process.

\subsubsection{Trust Builders in Video Selection (Before Watching)}
Before watching a video, participants often use comments to gauge user sentiment, build trust, and more easily select a suitable video. 
While video view count, creators, recommended videos, and thumbnails are often initial filters when selecting suitable videos, many participants rely on the comment section for further validation (details provided in Appendix~\ref{apx:exploratory-searching}). For instance, P8 shared, \squote{When deciding whether to watch or follow along with a video, I [first] gain trust from the comments.} P4 explained, \squote{I try to find positive comments when I [plan to] make a recipe. It helps in decision-making.} 
However, P9 noted how the lack of critical feedback hurts the reliability of videos.
During the co-watching session, P10 reviewed structured comments summarizing a recipe before starting a cooking video, and P6 read through comments, looking for both positive or critical comments, which influenced their decision to follow the video. This highlights how comments provide more nuanced evaluation, such as the effectiveness of exercise movements (P8), which are not always evident from the video itself. However, \squote[P11]{When comments ask about things too different [...] it might suggest a lack of trust,} in the video content. 

\subsubsection{Sensemaking and Personalized Learning (During Watching)}
Participants often turned to comments as a collaborative sensemaking tool to clarify and adapt video content, such as guidance on specific steps---especially in multi-step tasks such as recipes or hairstyling. 
For instance, comments that re-explain concepts were highlighted as particularly helpful, as P14 remarked, \squote{It is really helpful when the comments re-explain the content from the video using analogies [that] make it easier to understand.}
Some participants used comments for sensemaking about their own experience. During co-watching a hair-styling video, P11 explained, \squote{This comment says 'maybe it is not working because my hair is damaged,' maybe my hair is too damaged too.} 
In addition to explanations, some participants also highlighted the usefulness of summaries, timestamps, and step-specific explanations shared by others view (P2, P7, P8), which helped them navigate procedural tasks.
Creators also play an important role in collaborative sensemaking by responding to user queries. Several participants prioritize comments with validation from creators. For example, P4 explained \squote{In cases such as 'Can I make it using other ingredients? Can I use these substitutes?' in the comments. It is helpful if the video creator leaves a comment saying 'You can use this instead'.}
Similarly, during the co-watching session, participants tended to focus on comment threads with creator participation (P2, P13) or active discussion (P6, P14), expecting them to offer relevant explanations. 

Comments also supported personalized learning during watching. Participants used comments to tailor instruction to their own skill levels and situational constraints. As P11 explained, \squote{As a beginner at crocheting, I'm drawn to comments saying, 'as a beginner...', since I'm a beginner too.} Comments also help users adapt instructions when encountering unexpected variables or contextual constraints. For instance, in baking videos, P9 shared, \squote{Baking is particularly challenging for me because ingredient ratios are so important, and it's hard to make adjustments on my own. I try to follow the video as closely as possible, but when variability comes up, I check the comments more.} 
Additionally, some participants (P2, P6, P7, P14) browsed comments during less critical parts of a video, drawing on others’ reactions to stay engaged without losing track of the task.
For instance, during the co-watching session, P2 and P14 reviewed comments while the introduction was playing. P6, who was using an iPad, scrolled through the comments during playback, since \squote{seeing reactions from other people adds to my experience.}

\subsubsection{Extending and Reflecting on Learning (After Watching)}
After watching the video, participants often revisit the comment section to reflect on others' experiences, explore question threads, or seek updated information. 
P12 emphasized the importance of newer comments in software tutorials for staying current. Questions from other users can also address unanticipated gaps, as P6 shared, \squote{[User] questions allow me to get information on [information] I was curious about, but also [information] I hadn't thought of before.} 
Comments that contain critical feedback were noted as especially relevant to some participants for going beyond the provided knowledge (P1, P3, P8, P13). P1 explained, \squote{It's even more rewarding to see critical feedback from others. It's like getting a free lesson!} Similarly, during the co-watching session, P11 explained how \squote{the comments can be more helpful than the video itself.} 
Similarly, P6 used timestamped comments to revisit sections others found challenging in a video. When timestamps were missing, P13 turned on captions and scanned the timeline for words that match with the comment. 

\subsubsection{Challenges in Accessing Comments}
Participants surfaced many challenges they face when attempting to access or use comments. Although sorting options such as \textit{most liked} or \textit{most recent} are available, participants frequently relied on manual scrolling to locate relevant information, which was cumbersome due to spam and irrelevant content.
Locating relevant comments was particularly challenging when participants needed specific or domain-related information. For example, during a co-watching session, P14 turned to external resources when encountering domain-specific terminology.  
Accessing comments at the right moment also posed challenges. Navigating between video and comments often disrupted task flow, especially on mobile devices. As P7 remarked, \squote{If I leave the comment section, I have to find my place again, which is inconvenient.} These difficulties were further compounded during physical tasks, where hands-on engagement limited attention and interaction with comments. As P11 explained, \squote{My hands are busy with the task, so I probably won't look for comments now.}

\section{Design Opportunities for Future Work}
\label{sec:discussion}
Our findings show that while various challenges in accessing the comments exist, they are used throughout the task-learning process. Our participants used comments when searching for videos, making sense of the information, personalizing their learning, and reflecting or expanding beyond the knowledge contained in the video.
Based on these findings, we discuss design opportunities to better incorporate comments throughout the task-learning process as directions for future work.

\subsubsection{\textbf{Before Learning:} Video selection} 
Video selection is complex~\cite{fyfield2021NavigatingFour} and contains potential bias~\cite{ciampaglia2018HowAlgorithmic}. 
Our findings show that participants rely on comments to build trust and evaluate aspects of a video (e.g., estimating success). However, this evaluation currently depends on manual skimming and subjective interpretation. 
To better support informed video selection, collective signals embedded in comments could be surfaced more explicitly. For instance, sentiment analysis~\cite{bindhumol2024SentimentAnalysis,maw2024SentimentAnalysis} could visualize reported success and failure patterns, while summaries of influential comments (e.g., most liked or most replied to) may provide concise evaluative cues (\autoref{fig:teaser}-a).

%Video selection is complex~\cite{fyfield2021NavigatingFour} and contains potential bias~\cite{ciampaglia2018HowAlgorithmic}, comments can be used before the learning to aid selecting videos. Current comment use requires manual work in accessing each video and visually skimming the comments. Utilizing sentiment analysis~\cite{bindhumol2024SentimentAnalysis,maw2024SentimentAnalysis}, video comments could be visualized on a scale of reported success and failure counts by viewers (\autoref{fig:teaser}-a). Alternatively, word clouds~\cite{2020wordclouds} or summaries of most influential comments (e.g., most liked, most replied to), could be explored as means for users to objectively evaluate videos.

\subsubsection{\textbf{During Learning:} Personalized knowledge acquisition} 
Task-based learning often involves structured steps, where tacit knowledge plays a critical role~\cite{defelice2022LearningOthers}. Yet, this valuable knowledge is often underutilized or lost in video-learning environments~\cite{ramlogan2014ComparisonTwo,john2016CollationTwo}. 
Our findings show that participants use comments to clarify specific steps, access alternative explanations, and adapt instructions to their own skill level or situational constraints. 
Comments could be integrated into video playback to support collaborative sensemaking by surfacing re-explanations, common viewer questions, and additional user-provided knowledge.
%Comments could be integrated into video playback during the learning process to provide: additional information or knowledge, common viewer questions, or alternative explanations and analogies.
For instance, time anchored comments could be presented dynamically alongside the video using the current task-step as context. Similarly, inspired by OpinionSpace~\cite{faridani2010OpinionSpace}, comments can be visualized based on type (e.g., tips, questions), or communities based on known user characteristics---contained within the comments---during playback of each task step (\autoref{fig:teaser}-b).

\subsubsection{\textbf{After Learning:} Reflect through self and others' experience} 
Many participants interacted with the comments after the learning as well. Our findings show that comments can help reflect on different steps, and revisit steps with new acquired knowledge through the comments. 
While prior work explored ways to arrange comments chronologically and indicate sentiment~\cite{sung2016ToPINVisual}, it is aimed towards video creators. 
For viewers, implementing a step-based aggregation of comments can contribute to the reflection process. Grounding comments within task-steps, such as using a graph-based approach inspired by \citet{oomori2024SkillsInterpreterCase}, directly contributes to task-learning. For instance, viewers can explore steps with high user interaction, or view comments relevant to a step they find challenging (\autoref{fig:teaser}-c). 

\subsubsection{\textbf{Improving Access:} Smaller screens and visual accessibility} 
Beyond the learning process, our findings reveal an overall lack of access to comments, in particular on mobile devices. 
With the rise of mobile-first, short-form platforms~\cite{nguyen2023HowCivil,violot2024ShortsVs}, ensuring comment accessibility on smaller screens is increasingly important. Tools like \citet{kim2022FitVidResponsive} adapt video visuals for mobile; comments require similar attention~\cite{manca2018SocialNetwork,towne2017ConflictComments,lee2017MakingSense}.
While not part of this study, such barriers can affect underserved groups, such as blind and low-vision (BLV) users.
Our findings show that current practices require users to visually skim through comments throughout all stages of the learning, a challenging task for BLV users~\cite{sechayk2025veasyguide}.
DanmuA11y~\cite{xu2025danmua11y} explored non-visual access to time-anchored comments; however, they do not support user-lead exploration of comments. 
Increasing the accessibility of comments is a promising yet understudied area~\cite{chen2024InclusiveVideo}.

\section{Study Limitations}
Our study focuses primarily on the user experience of consuming comments, as none of the participants reported regularly writing comments. This limits the scope of our findings to comment consumption. While prior work has analyzed user commenting patterns~\cite{du2009FeltMore}, much of it relies on existing comment data~\cite{wang2023LetsPlay,thelwall2012CommentingYouTube,benson2015CommentingLearn}. Future research should investigate the motivations behind user commenting behavior to provide a more comprehensive understanding.
Additionally, our study's participant pool is limited in size and cultural diversity. 
As most participants were university-affiliated young adults, the findings may not fully generalize to broader populations~\cite{sung2018SearchingVideos}. Furthermore, since YouTube was the only platform used in this study, our findings may not generalize to other platforms with different comment systems. Future work should examine more diverse populations and platform contexts to better understand variations in comment use and user experience.

\section{Conclusion}
\label{sec:conclusion}
Social interactions are valuable for learning, including online video-based learning. In task-learning specifically, the lack of viewer agency in videos limits the ability to acquire tacit knowledge and adapt the learning process. To address this challenge and leverage the abundance of online task-learning videos, many users turn to the comment section. Our exploratory user evaluation uncovers how comments are utilized throughout the learning process: before, during, and after the viewing. We discuss further opportunities to enhance task-learning through features like context-aware comments, alternative visualizations, and improved accessibility. We believe that our results encourage a reconsideration of the comment section as a collaborative learning infrastructure that extends beyond task-learning.

\begin{acks}
    We are grateful to the anonymous reviewers for their insightful feedback, and to all participants whose contributions enriched this study.
Nayoung Kim was supported by the DGIST Undergraduate Research Award (DURA) Program of Daegu Gyeongbuk Institute of Science and Technology (DGIST).    
Yotam Sechayk was supported by the Ministry of Education, Culture, Sports, Science and Technology (MEXT) Monbukagakusho Scholarship.
This work was partially supported by Japan Science and Technology Agency (JST) as part of Adopting Sustainable Partnerships for Innovative Research Ecosystem (ASPIRE), Grant Number JPMJAP2401.
\end{acks}

\bibliographystyle{ACM-Reference-Format}
\bibliography{ref.global,ref.main}

\appendix
% \section{Platform Analysis Details}
% \label{apx:platform}

% \begin{table*}[h]
% \centering
% \caption{Summary of Reader related interaction features of platforms. (*) Only YouTube and NicoNico allow for in-time display of comment, with only \textit{YouTube Live} videos available. (**) User rating is not available per individual video.}
% \input{tables/05c-platform-reading}
% \label{tab:platform-reading-summary}
% \end{table*}

% \begin{table*}[h]
% \centering
% \caption{Summary of Writing related interaction features of platforms.}
% \input{tables/05d-platform-writing}
% \label{tab:platform-writing-summary}
% \end{table*}

\section{Exploratory Study Interview Questions}
\label{apx:exploratory-questions}
\subsection{Demographic}
    \begin{enumerate}
        \item What is your age?
        \item What is your gender identity?
    \end{enumerate}
\subsection{Task-learning Video Watching Experience}
\begin{enumerate}
    \item How frequently do you learn new skills with task-learning videos?
    \item What tasks do you usually learn with videos?
    \item Are there any differences between learning with videos compared with written media?
    \item How do you usually find task-learning videos you want to watch?
    \item Do you have a specific device you prefer to use for task-learning?
\end{enumerate}
\subsection{Comment Usage Habits}
\begin{enumerate}
    \item Do you read or write comments during task-learning with videos?
    \item How do you usually navigate the comment section?
    \item Do you think that there are situations where you navigate to the comment section more or less?
    \item Do you think that some comments are more helpful than others?
    \item Are you likely to comment on videos yourself?
    \item Are there any difficulties you experience about reading or writing comments?
\end{enumerate}

\section{Video Searching Strategies}
\label{apx:exploratory-searching}
\begin{figure}[h]
    \centering
    \includegraphics[width=\linewidth]{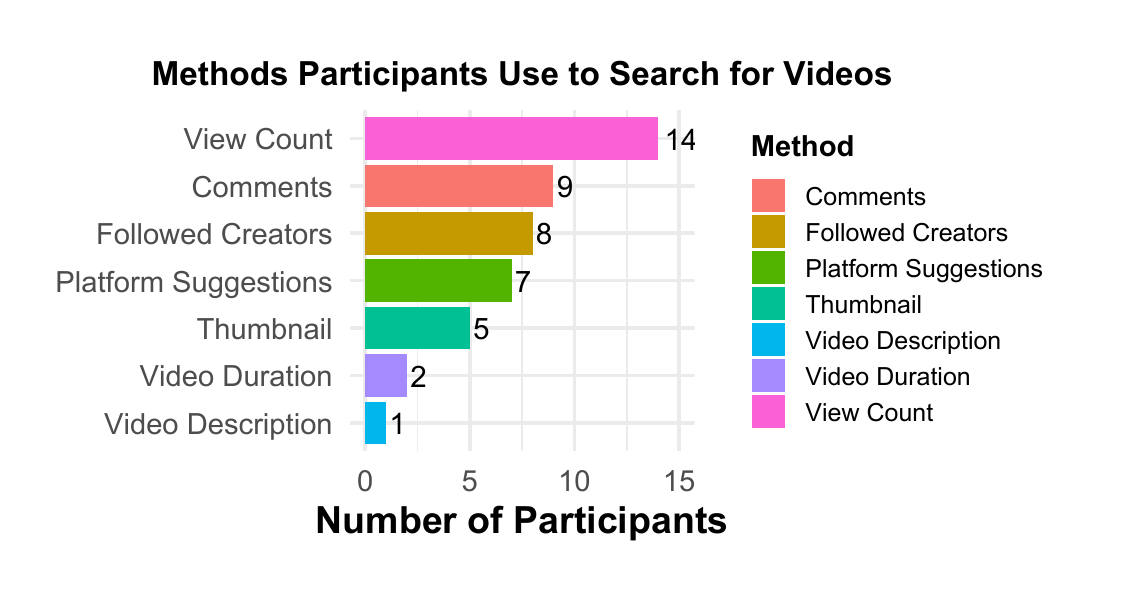}
    \caption{Number of participants for using different approaches when searching for suitable videos in task-learning scenarios. In total there are 14 participants.}
    \Description{The figure is a horizontal bar chart titled "Methods Participants Use to Search for Videos." It features different colored bars representing the number of participants using various methods to search for videos. The methods, which are listed on the left, include View Count, Comments, Followed Creators, Platform Suggestions, Thumbnail, Video Duration, and Video Description. Each method is associated with a specific color in the chart's legend on the right. View Count is the most popular method, with 14 participants using it, followed by Comments with 9, Followed Creators with 8, Platform Suggestions with 7, Thumbnail with 5, Video Duration with 2, and Video Description with 1. The x-axis is labeled "Number of Participants," and it extends from 0 to 15.}
    \label{fig:enter-label}
\end{figure}

\newpage

\section{Ethical Considerations}
\label{apx:ethical}
Participation in this study was voluntary, and all participants provided informed consent prior to their involvement. Participants were informed of their right to skip any question during the interviews or co-watching sessions without consequence. For the co-watching sessions, we relied on existing content moderation mechanisms and filtering practices implemented by the platform (YouTube) to ensure the absence of harmful content. Participants were free to select their own subjects and videos, and they could navigate between videos as they wished.
All participant data was anonymized, and user identities were replaced with codes (e.g., P1) during analysis. Recordings of the interviews and co-watching sessions were securely stored and were discarded following the completion of the analysis. Ethical guidelines were followed throughout the study, ensuring confidentiality, participant autonomy, and compliance with data protection standards.
\end{document}